\documentclass{article}

\usepackage{PRIMEarxiv}

\usepackage[utf8]{inputenc} 
\usepackage[T1]{fontenc}    
\usepackage{hyperref}       
\usepackage{url}            
\usepackage{booktabs}       
\usepackage{amsfonts}       
\usepackage{nicefrac}       
\usepackage{microtype}      
\usepackage{lipsum}
\usepackage{fancyhdr}       
\usepackage{graphicx}       
\usepackage{orcidlink}
\graphicspath{{media/}}     

\usepackage{pgf-pie} 
\usepackage{xcolor} 

\usepackage{tikz}
\usetikzlibrary{shapes.geometric, arrows, positioning}
\usetikzlibrary{shadows}

\title{AI-Powered Assistive Technologies for Visual Impairment}

\author{
  Prudhvi Naayini \orcidlink{0009-0009-5010-2092}, Praveen Kumar Myakala \orcidlink{0009-0009-5010-2092} and Chiranjeevi Bura \orcidlink{0009-0001-1223-300X}\\
  University of Colorado Boulder \\
  Boulder, CO 80309 USA \\
  \texttt{\{Prudhvi.Naayini, Praveen.Myakala, Chiranjeevi.Bua\}@colorado.edu}
  \And
  Anil Kumar Jonnalagadda \orcidlink{0009-0000-8207-4131}\\
  Senior Data Engineer, Google LLC \\ 
  Mountain View, CA 94043, USA
  \And
  Srikanth Kamatala \orcidlink{0009-0000-2375-7119}\\
  Director Business Intelligence, Klein Tools \\ 
  Mansfield, TX 76063, USA\\
}

\begin{document}
\maketitle

\begin{abstract}
Artificial Intelligence (AI) is revolutionizing assistive technologies. It offers innovative solutions to enhance the quality of life for individuals with visual impairments. This review examines the development, applications, and impact of AI-powered tools in key domains, such as computer vision, natural language processing (NLP), and wearable devices. Specific advancements include object recognition for identifying everyday items, scene description for understanding surroundings, and NLP-driven text-to-speech systems for accessing digital information. Assistive technologies like smart glasses, smartphone applications, and AI-enabled navigation aids are discussed, demonstrating their ability to support independent travel, facilitate social interaction, and increase access to education and employment opportunities.

The integration of deep learning models, multimodal interfaces, and real-time data processing has transformed the functionality and usability of these tools, fostering inclusivity and empowerment. This article also addresses critical challenges, including ethical considerations, affordability, and adaptability in diverse environments. Future directions highlight the need for interdisciplinary collaboration to refine these technologies, ensuring equitable access and sustainable innovation. By providing a comprehensive overview, this review underscores AI's transformative potential in promoting independence, enhancing accessibility, and fostering social inclusion for visually impaired individuals.
\end{abstract}


\section{Introduction}

Visual impairment is a significant global challenge, affecting over 2.2 billion individuals worldwide, with at least 1 billion cases preventable or treatable \cite{who_visual_impairment}. Beyond personal hardship, visual impairment imposes substantial societal costs, as highlighted by the global economic burden of blindness, which includes lost productivity and healthcare expenses \cite{smith1996economic}. These challenges underscore the urgent need for innovative solutions to enhance accessibility and autonomy for affected individuals.  

Artificial intelligence (AI) is proving transformative in addressing accessibility challenges, particularly for the visually impaired. With advancements in machine learning and deep learning, AI-powered systems can perform complex tasks such as real-time object recognition, scene analysis, and natural language processing (NLP) with unprecedented accuracy and efficiency \cite{seeing_ai, envision_glasses_review, reuters_robot_dog}.  

For instance, AI-powered applications such as Microsoft's Seeing AI utilize computer vision and NLP to identify objects, text, people, and scenes, offering audio descriptions that enhance users' understanding of their environment \cite{seeing_ai}. Similarly, wearable devices like Envision Glasses combine sophisticated cameras with real-time speech synthesis to deliver contextual information, enabling visually impaired users to interact more effectively with their surroundings \cite{envision_glasses_review}. Additionally, experimental technologies such as robotic guide dogs aim to further enhance mobility and independence for users, incorporating advanced sensors and AI-driven navigation systems \cite{reuters_robot_dog}. These capabilities have the potential to significantly improve the quality of life for visually impaired individuals, fostering greater independence and well-being.  

The purpose of this review is to explore the landscape of AI-powered assistive technologies for visual impairment. Specifically, we examine the development and applications of tools leveraging computer vision, NLP, and wearable devices. We discuss their impact on daily living, mobility, education, and social interaction, highlighting their potential to transform accessibility for individuals with visual impairments.  

This paper is organized as follows: Section~\ref{sec:background} provides an overview of visual impairment and traditional assistive technologies. Section~\ref{sec:ai_technologies} focuses on AI-powered tools, including computer vision, natural language processing (NLP), and wearable devices. Section~\ref{sec:applications} explores real-world applications and their impacts on areas such as independent living, mobility, and education. Section~\ref{sec:challenges} addresses key challenges, including issues of affordability, accessibility, and ethical considerations. Section~\ref{sec:future_directions} outlines potential future directions, highlighting innovations, interdisciplinary collaborations, and strategies for enhancing inclusivity and usability. Finally, Section~\ref{sec:conclusion} summarizes the key findings and emphasizes the importance of addressing challenges while leveraging opportunities for advancement.

\section{Background}
\label{sec:background}

Visual impairment encompasses a wide range of conditions, from partial vision loss to complete blindness. The World Health Organization (WHO) classifies visual impairment into four categories: mild, moderate, severe, and blindness \cite{who_visual_impairment}. Common causes include uncorrected refractive errors, cataracts, glaucoma, diabetic retinopathy, and macular degeneration.  

To better understand the prevalence and causes of visual impairment, Table~\ref{table:prevalence} summarizes global data on the distribution of these conditions.

\begin{table}[h!]
 \caption{Global Prevalence of Visual Impairment by Cause (2023)}
  \centering
  \begin{tabular}{lll}
    \toprule
    Cause                        & Prevalence (\%) & Treatability                                     \\
    \midrule
    Uncorrected refractive errors & 49\%            & High (Glasses, Contact Lenses)                  \\
    Cataracts                     & 23\%            & High (Cataract Surgery)                         \\
    Glaucoma                      & 10\%            & Moderate (Eye Drops, Laser Treatment, Surgery)  \\
    Diabetic Retinopathy          & 6\%             & Moderate (Laser Treatment, Injections, Surgery) \\
    Macular Degeneration          & 12\%            & Low (Limited Therapy or Surgery)                \\
    \bottomrule
  \end{tabular}
  \label{table:prevalence}
\end{table}

\begin{figure}[h!]
    \centering
    \begin{tikzpicture}
        \pie[text=legend, radius=2.5, color={
            cyan!80, orange!80, gray!60, teal!80, magenta!80
        }, drop shadow]{
            49/Uncorrected Refractive Errors,
            23/Cataracts,
            10/Glaucoma,
            6/Diabetic Retinopathy,
            12/Macular Degeneration
        }
    \end{tikzpicture}
    \caption{Distribution of Visual Impairment by Cause (2023)}
    \label{fig:causes_pie_chart}
\end{figure}
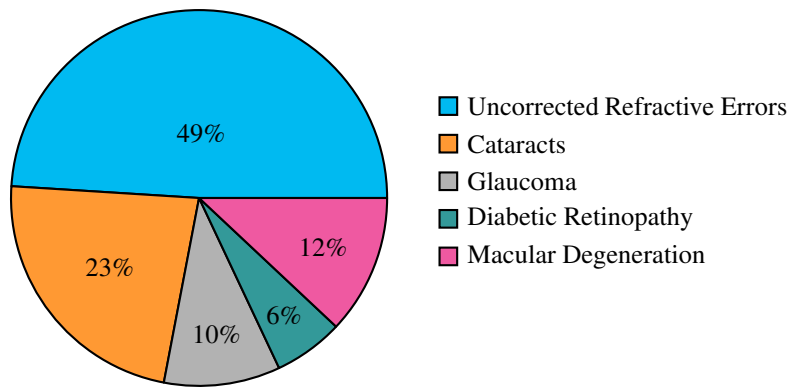

This data highlights that a significant portion of visual impairment is preventable or treatable, with cataracts and uncorrected refractive errors being the most addressable. Figure~\ref{fig:causes_pie_chart} visually illustrates the distribution of these causes, emphasizing the opportunity for effective interventions.

\subsection{Traditional Assistive Technologies}

Historically, assistive technologies for the visually impaired have relied on non-digital or rudimentary electronic solutions. Table~\ref{table:traditional_vs_ai} compares traditional assistive tools with modern AI-powered alternatives, highlighting the evolution of technology in this domain.

\begin{table}[h!]
 \caption{Comparison of Traditional and AI-Powered Assistive Technologies}
  \centering
  \begin{tabular}{lll}
    \toprule
    Category             & Traditional Technology          & AI-Powered Alternative                        \\
    \midrule
    Mobility             & White Cane                     & AI-enabled Navigation Systems                \\
    Text Access          & Braille Books                  & Screen Readers, Text-to-Speech               \\
    Object Identification & Manual Identification         & Real-time Object Recognition                 \\
    Social Interaction    & None                          & AI-based Facial Recognition                  \\
    Information Access    & Radio, Audiobooks             & Personalized News Feeds, Podcasts            \\
    Environmental Awareness & Limited                     & Smart Home Integration, Obstacle Detection   \\
    \bottomrule
  \end{tabular}
  \label{table:traditional_vs_ai}
\end{table}

Despite their utility, traditional tools often lack adaptability, personalization, and advanced features. These limitations have driven the development of AI-powered alternatives, which offer enhanced functionality, real-time adaptability, and seamless integration into users’ daily lives \cite{techstrong_ai}.
 


\subsection{Mathematical Framework for Accessibility Metrics}

AI-powered technologies often rely on quantitative metrics to evaluate their effectiveness. For example, accessibility metrics \( A \) for assistive technologies can be defined as:

\[
A = \frac{U_f}{U_t} \times 100
\]

where:
\begin{itemize}
    \item \( U_f \): Number of functionalities accessible to visually impaired users.
    \item \( U_t \): Total number of functionalities available to fully sighted users.
\end{itemize}

This equation ensures that accessibility improvements can be measured quantitatively. For example, recent studies have applied similar frameworks to evaluate the effectiveness of wearable assistive devices and AI-enabled navigation systems \cite{lin2019wearable_assistive_system}.

With the limitations of traditional assistive technologies in mind, we now delve into the advancements brought about by artificial intelligence in the next section.

\section{AI-Powered Assistive Technologies}
\label{sec:ai_technologies}

Recent advancements in artificial intelligence (AI) have enabled the development of sophisticated assistive technologies tailored to the needs of visually impaired individuals. This section explores key domains where AI is transforming accessibility: computer vision, natural language processing (NLP), and wearable devices.

\subsection{Computer Vision}

Computer vision systems empower visually impaired users by converting visual data into accessible formats. These technologies rely on advanced deep learning models for real-time interpretation of the surrounding environment. Table~\ref{table:computer_vision_apps} highlights popular AI-powered applications that leverage computer vision.

\begin{table}[h!]
 \caption{AI-Powered Computer Vision Applications for Visual Impairment}
  \centering
  \begin{tabular}{lll}
    \toprule
    \textbf{Application}  & \textbf{Functionality}                          & \textbf{Key Features}                 \\
    \midrule
    Seeing AI             & Real-time object and text recognition         & Scene descriptions, barcode scanning  \\
    Google Lookout        & Scene and object detection                    & Currency identification, text reading \\
    Be My Eyes            & Volunteer-assisted video interpretation       & Real-time assistance from volunteers  \\
    \bottomrule
  \end{tabular}
  \label{table:computer_vision_apps}
\end{table}

For instance, Microsoft's \textbf{Seeing AI} app enables users to identify products in a grocery store by reading barcodes, while also providing detailed scene descriptions to assist in navigation. Similarly, \textbf{Google Lookout} helps users navigate indoor spaces, such as identifying doors, furniture, or objects in a cluttered room. \textbf{Be My Eyes} connects users with sighted volunteers through a live video call for assistance with more complex tasks.

\tikzstyle{startstop} = [ellipse, rounded corners, minimum width=2.5cm, minimum height=1cm, text centered, draw=black, fill=cyan!50, drop shadow]
\tikzstyle{process} = [rectangle, minimum width=3cm, minimum height=1cm, text centered, draw=black, fill=orange!50, drop shadow]
\tikzstyle{arrow} = [thick,->,>=stealth, drop shadow]
\tikzstyle{textbelow} = [align=center, font=\scriptsize] 

\begin{figure}[h!]
    \centering
    \begin{tikzpicture}[node distance=3.5cm] 

        \node (input) [startstop] {Input};
        \node[textbelow, below=0.25cm of input] {Raw Visual Data \\ (Camera/Sensors)};

        \node (preprocessing) [process, right of=input] {Preprocessing};
        \node[textbelow, below=0.25cm of preprocessing] {Noise Reduction, \\ Image Enhancement};

        \node (feature) [process, right of=preprocessing] {Feature Extraction};
        \node[textbelow, below=0.25cm of feature] {Key Features \\ (Edges, Objects, Text)};

        \node (inference) [process, right of=feature] {Inference};
        \node[textbelow, below=0.25cm of inference] {Object Detection, \\ Scene Analysis};

        \node (output) [startstop, right of=inference] {Output};
        \node[textbelow, below=0.25cm of output] {Accessible Formats \\ (Speech, Braille, Haptic Feedback)};

        \draw [arrow] (input) -- (preprocessing);
        \draw [arrow] (preprocessing) -- (feature);
        \draw [arrow] (feature) -- (inference);
        \draw [arrow] (inference) -- (output);

    \end{tikzpicture}
    \caption{Computer Vision Pipeline for Assistive Applications}
    \label{fig:cv_diagram}
\end{figure}
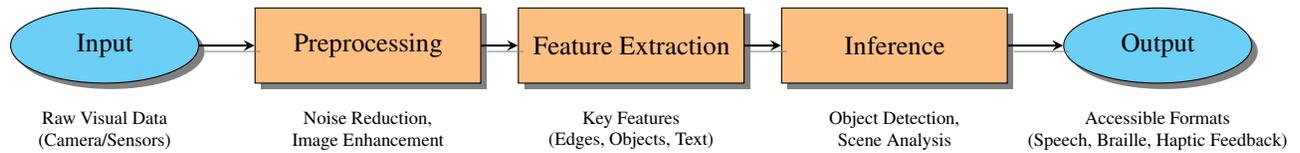

While these tools offer immense benefits, they are not without limitations. For example, object recognition may fail in poor lighting conditions, and some apps require constant internet connectivity, which can be a challenge in remote or low-resource areas \cite{seeing_ai, google2023lookout}.

Beyond visual perception, natural language processing (NLP) plays a crucial role in enabling communication and information access for visually impaired individuals.

\subsection{Natural Language Processing (NLP)}

Natural Language Processing (NLP) technologies are foundational in enabling visually impaired individuals to communicate, interact, and access digital content. Tools such as screen readers, text-to-speech (TTS) systems, and voice assistants make information more accessible. Key applications are summarized in Table~\ref{table:nlp_apps}.

\begin{table}[h!]
 \caption{AI-Powered NLP Applications for Visual Impairment}
  \centering
  \begin{tabular}{lll}
    \toprule
    \textbf{Application}    & \textbf{Functionality}           & \textbf{Key Features}                    \\
    \midrule
    JAWS Screen Reader      & Text-to-speech conversion       & Braille output, multi-language support   \\
    NaturalReader           & Document and web page reading  & AI-enhanced natural voice synthesis      \\
    Siri / Alexa            & Voice commands and interaction & Hands-free device control, information retrieval \\
    \bottomrule
  \end{tabular}
  \label{table:nlp_apps}
\end{table}

For example, students can use \textbf{JAWS Screen Reader} to read and navigate textbooks or digital documents, including mathematical symbols and graphs, with Braille output. Meanwhile, \textbf{NaturalReader} provides AI-enhanced voice synthesis for seamless reading of web pages or PDFs. Voice assistants like \textbf{Siri} and \textbf{Alexa} allow users to perform tasks such as setting reminders, retrieving weather updates, or controlling smart home devices—all hands-free.

\tikzstyle{main} = [rectangle, rounded corners, minimum width=4cm, minimum height=2cm, text centered, draw=black, fill=cyan!10, drop shadow]
\tikzstyle{sub} = [rectangle, minimum width=4cm, minimum height=1cm, text centered, draw=black, fill=orange!60, drop shadow]
\tikzstyle{io} = [rectangle, minimum width=3cm, minimum height=1cm, text centered, draw=black, fill=cyan!50, drop shadow]
\tikzstyle{arrow} = [thick, ->, >=stealth]

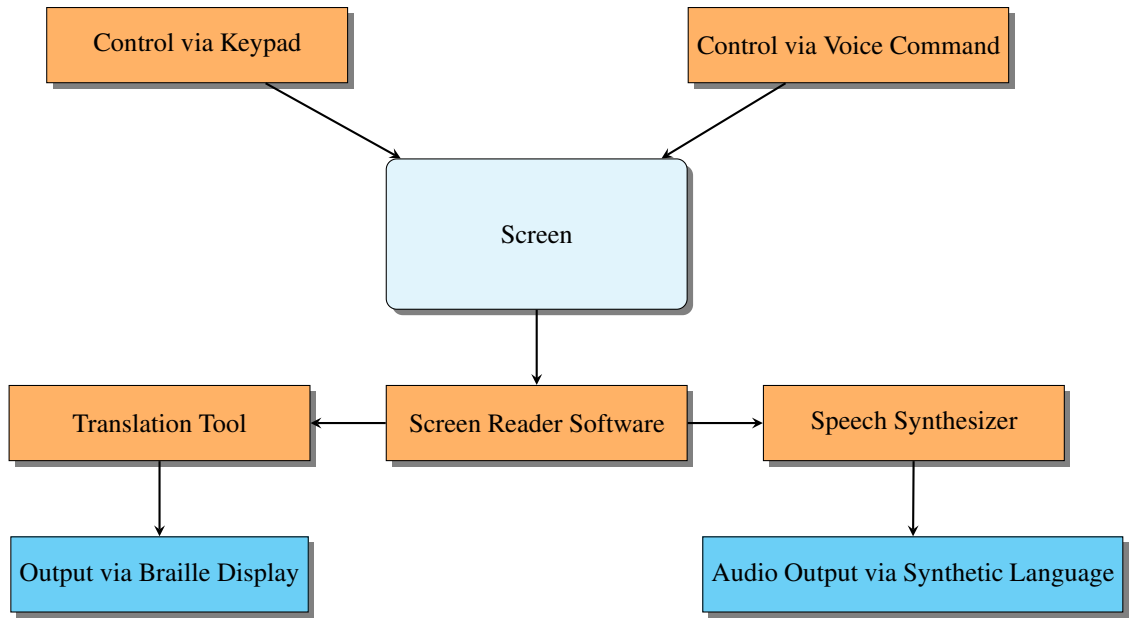
\begin{figure}[h!]
    \centering
    \begin{tikzpicture}[node distance=0.5cm and 0.5cm]

        \node (screen) [main] {Screen};

        \node (keypad) [sub, above left=of screen, xshift=-0cm, yshift=0.5cm] {Control via Keypad};
        \node (voice) [sub, above right=of screen, xshift=-0.5cm, yshift=0.5cm] {Control via Voice Command};

        \node (screenreader) [sub, below=2cm of screen, yshift=1cm] {Screen Reader Software};

        \node (translation) [sub, left=3cm of screenreader, xshift=2cm] {Translation Tool};

        \node (braille) [io, below=2cm of translation, yshift=1cm] {Output via Braille Display};
        \node (audio) [io, below=2cm of screenreader, xshift=5cm, yshift=1cm] {Audio Output via Synthetic Language};

        \node (synthesizer) [sub, right=3cm of screenreader, xshift=-2cm] {Speech Synthesizer};

        \draw [arrow] (keypad) -- (screen);
        \draw [arrow] (voice) -- (screen);

        \draw [arrow] (screen) -- (screenreader);
        \draw [arrow] (screenreader) -- (translation);
        \draw [arrow] (translation) -- (braille);
        \draw [arrow] (screenreader) -- (synthesizer);
        \draw [arrow] (synthesizer) -- (audio);

    \end{tikzpicture}
    \caption{Flowchart for Assistive Technology: Screen Interaction and Outputs}
    \label{fig:assistive_flowchart}
\end{figure}

Despite their widespread use, these tools face limitations such as difficulty in handling highly technical or ambiguous content and occasional errors in multi-accent environments. Similarly, NLP systems may struggle with uncommon languages or dialects, making accessibility more challenging for certain user groups \cite{gavat2023natural}.

\subsection{Wearable Devices}

Wearable assistive devices combine AI technologies with portability to provide real-time support for visually impaired individuals. These devices range from smart glasses to haptic feedback systems. Table~\ref{table:wearable_devices} lists popular wearable technologies.

\begin{table}[h!]
 \caption{AI-Powered Wearable Devices for Visual Impairment}
  \centering
  \begin{tabular}{lll}
    \toprule
    \textbf{Device}         & \textbf{Functionality}               & \textbf{Key Features}                     \\
    \midrule
    Envision Glasses        & Real-time object and text detection & Speech synthesis, navigation assistance   \\
    OrCam MyEye             & Wearable AI-powered camera          & Facial recognition, text-to-speech        \\
    Sunu Band               & Haptic navigation system            & Ultrasonic obstacle detection, vibrations \\
    \bottomrule
  \end{tabular}
  \label{table:wearable_devices}
\end{table}

For example, \textbf{Envision Glasses} integrate a camera and speech synthesis to help users identify text on signage, read printed materials, or recognize objects in their environment. \textbf{OrCam MyEye} uses a wearable camera to perform tasks such as facial recognition, allowing users to identify friends and family members. The \textbf{Sunu Band} provides tactile feedback to guide users around obstacles using ultrasonic sensors, making it a valuable tool for independent mobility.

However, wearable devices also come with limitations. Battery life is a significant concern, as these devices require frequent recharging to ensure functionality. Additionally, affordability remains a challenge, as some devices are prohibitively expensive for users in low-income settings \cite{brilli2024airis}.

\subsection{Summary of AI-Powered Technologies}

AI-powered assistive technologies significantly enhance the independence and quality of life for visually impaired individuals. Table~\ref{tab:ai_technologies_summary} provides a summary of the key technologies discussed.

\begin{table}[h!]
    \centering
    \caption{Summary of AI-Powered Assistive Technologies}
    \begin{tabular}{lll}
        \toprule
        \textbf{Technology}            & \textbf{Functionality}                         & \textbf{Key Benefits}                              \\
        \midrule
        Smart Glasses                  & Object recognition, text reading, facial recognition & Promotes independence in daily tasks             \\
        Voice Assistants               & Hands-free control of devices, task automation   & Enhances accessibility and productivity           \\
        Navigation Aids                & Real-time obstacle detection, scene description & Facilitates independent mobility                  \\
        Screen Readers                 & Text-to-speech for digital content              & Provides access to education and workplace tools  \\
        Braille Displays               & Tactile access to text and symbols              & Enables inclusive education, supports coding      \\
        Facial Recognition Tools       & Identifies people in social settings            & Improves confidence in social interactions        \\
        Assistance Apps                & Volunteer-based remote assistance               & Offers flexible, on-demand support                \\
        \bottomrule
    \end{tabular}
    \label{tab:ai_technologies_summary}
\end{table}

The following sections explore the real-world applications of these technologies and their impact on daily living, mobility, education, and social interaction.

\section{Applications and Impact of Assistive Technologies}
\label{sec:applications}

AI-powered assistive technologies have a profound impact on the daily lives of individuals with visual impairments. This section explores their applications in independent living, navigation, education, social interaction, and mental health, highlighting the transformative potential of these tools.

\subsection{Independent Living}

AI-enabled devices play a crucial role in promoting independence in daily activities. This subsection focuses on two key innovations: \textbf{smart glasses} and \textbf{voice assistants}.

\subsubsection{Smart Glasses}

Smart glasses such as Envision Glasses and OrCam MyEye provide real-time object recognition, text reading, and facial recognition, enabling users to navigate their environments and perform tasks without assistance \cite{orcam_2023, envision_glasses_review}. For example, these devices allow users to identify household items, read labels on medication bottles, and locate objects in cluttered spaces.


\subsubsection{Voice Assistants}

Voice assistants such as Alexa and Siri enable hands-free control of smart home devices, such as adjusting thermostats, turning lights on and off, and managing appliances. These tools foster greater autonomy and reduce reliance on caregivers \cite{voice_assistants_review}. For instance, a user can ask Alexa to read the news aloud or control their smart thermostat via voice commands.

\subsection{Navigation and Mobility}

Navigation is one of the most critical challenges faced by visually impaired individuals. AI-powered navigation aids, such as the \textbf{Sunu Band} and smartphone applications like \textbf{Seeing AI}, address this issue by providing real-time guidance.

\begin{itemize}
    \item \textbf{Sunu Band:} This wearable device uses ultrasonic sensors to detect obstacles and provides haptic feedback to guide users safely through their surroundings \cite{sunu_navigation_study}.
    \item \textbf{Seeing AI:} This smartphone app offers features such as text-to-speech conversion, object detection, and scene description, allowing users to understand their environment better \cite{seeing_ai}.
\end{itemize}

These technologies enhance mobility, enabling visually impaired individuals to travel independently and confidently. Beyond enhancing mobility, AI-powered assistive technologies also play a crucial role in improving access to education and employment opportunities.

\subsection{Education and Employment}

Access to education and employment opportunities has been significantly enhanced through AI-powered assistive technologies. This subsection highlights tools like \textbf{screen readers}, \textbf{Braille displays}, and \textbf{voice assistants} that contribute to inclusivity in education and workplaces.

\subsubsection{Screen Readers}

Screen readers, such as JAWS and NVDA, allow students and professionals to access digital content, including books, research articles, and workplace documents \cite{freedom_scientific_jaws,nv_access_nvda}. These tools provide multi-language support and are essential for navigating online platforms.

\subsubsection{Braille Displays}

AI-enhanced Braille displays provide tactile access to textual materials, making education more inclusive \cite{microsoft_accessibility_math}. These displays are especially useful for subjects like mathematics and programming, where visual diagrams and symbols are difficult to interpret audibly.

\subsubsection{Voice Assistants}

In addition to screen readers and Braille displays, voice assistants contribute to professional environments by enabling task automation, scheduling meetings, and setting reminders. These tools ensure that visually impaired individuals remain competitive in the workplace \cite{voice_assistants_productivity}.

\begin{table}[h!]
 \caption{AI Tools Supporting Education and Employment}
  \centering
  \begin{tabular}{lll}
    \toprule
    \textbf{Tool}             & \textbf{Application}              & \textbf{Key Features}                   \\
    \midrule
    JAWS Screen Reader        & Reading digital content           & Multi-language support, Braille output \\
    NVDA                      & Web navigation and document access & Free and open-source                   \\
    AI-Enhanced Braille Displays & Tactile reading and writing       & Supports mathematics and code           \\
    \bottomrule
  \end{tabular}
  \label{tab:education_employment}
\end{table}

\subsection{Social Interaction}

Social interaction is often a challenging area for visually impaired individuals due to the lack of access to non-verbal cues such as facial expressions and body language. AI-powered tools such as \textbf{facial recognition smart glasses} and \textbf{assistance apps} help bridge this gap.

\subsubsection{Facial Recognition Smart Glasses}

Tools like OrCam MyEye allow users to identify people in social settings through facial recognition \cite{orcam_myeye}. For example, a user can receive an audio prompt identifying a friend's name during a conversation.

\subsubsection{Assistance Apps}

Applications like Be My Eyes connect visually impaired users with sighted volunteers through live video calls, enabling real-time assistance with tasks such as choosing clothing or identifying items in unfamiliar environments \cite{be_my_eyes_data}.

\begin{figure}[h!]
    \centering
    \begin{tikzpicture}

    \draw[thick] (0,0) circle(1cm) node {\textbf{User}};
    \draw[thick, red] (-1,0.5) -- (1,0.5); 
    \draw[thick, red] (-1,0.5) -- (-0.5,0.8);
    \draw[thick, red] (1,0.5) -- (0.5,0.8);

    \draw[thick, fill=cyan!60] (4,0) circle(1cm) node {\textbf{Friend}};
    
    \draw[thick, green!50!black] (4,1.2) rectangle ++(2,0.6);
    \node at (5,1.5) {Name: John};

    \draw[->, thick, blue] (1,0) -- (3,0)
        node[midway, above] {Recognition};

    \draw[gray, dashed] (-2,-1.5) rectangle (6,2.5); 
    \node[below right] at (-2,2.5) {Social Setting};
    \end{tikzpicture}
    
    \caption{User identifying a friend in a social setting using smart glasses}
    \label{fig:social_interaction}
\end{figure}
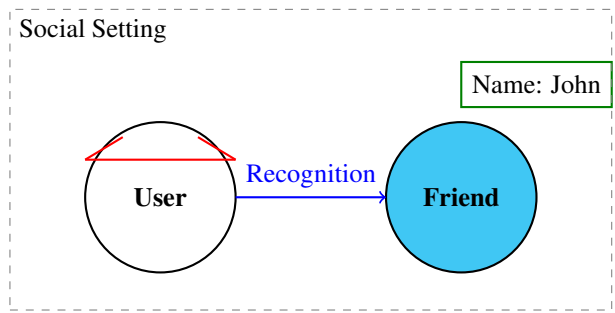

By breaking down communication barriers, these tools enhance social inclusion and improve the quality of life for visually impaired individuals.

\subsection{Mental Health and Well-being}

The autonomy and accessibility provided by AI-powered technologies contribute to improved mental health and overall well-being. Reducing dependence on others fosters a sense of independence and boosts self-esteem. However, the lack of accessibility or affordability of these tools may negatively impact mental health, contributing to feelings of frustration, exclusion, and helplessness \cite{ai_mental_health_assessment,ai_assistive_technology_wellbeing}.

    \tikzstyle{startstop} = [rectangle, rounded corners, minimum width=3cm, minimum height=1cm,text centered, draw=black, fill=cyan!60, drop shadow]
    \tikzstyle{process} = [rectangle, minimum width=3cm, minimum height=1cm, text centered, draw=black, fill=orange!60, drop shadow]
    \tikzstyle{arrow} = [thick,->,>=stealth]

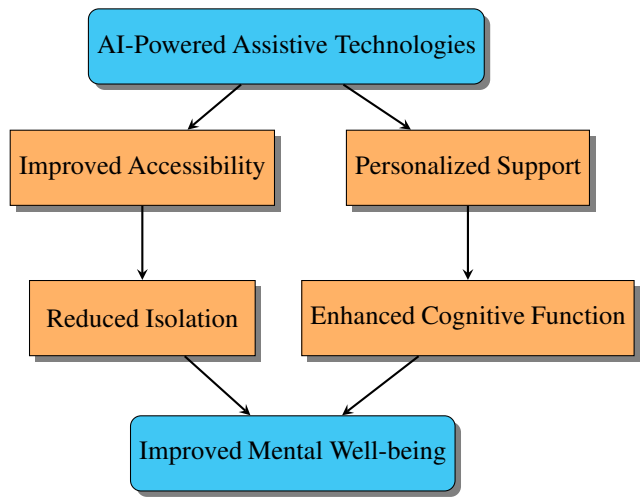
\begin{figure}[h!]
    \centering
    \begin{tikzpicture}[node distance=2cm]

    \node (start) [startstop] {AI-Powered Assistive Technologies};
    \node (process1) [process, below left of=start,xshift=-0.5cm, yshift=-0.2cm] {Improved Accessibility};
    \node (process2) [process, below right of=start,xshift=1cm, yshift=-0.2cm] {Personalized Support};
    \node (process3) [process, below of=process1,xshift=-0cm, yshift=0 cm] {Reduced Isolation};
    \node (process4) [process, below of=process2] {Enhanced Cognitive Function};
    \node (end) [startstop, below of=process3,xshift=2.0cm, yshift=0.2cm] {Improved Mental Well-being};
    
    \draw [arrow] (start) -- (process1);
    \draw [arrow] (start) -- (process2);
    \draw [arrow] (process1) -- (process3);
    \draw [arrow] (process2) -- (process4);
    \draw [arrow] (process3) -- (end);
    \draw [arrow] (process4) -- (end);
    
    \end{tikzpicture}

    \caption{Impact of AI-powered assistive technologies on mental health and well-being.}
    \label{fig:mental_health}
\end{figure}

\subsection{Policy Implications}

While the benefits of AI-powered assistive technologies are significant, their widespread adoption depends on affordability, accessibility, and supportive policies. For instance, initiatives such as government subsidies for assistive technologies in countries like Norway and Australia have successfully increased adoption rates \cite{norway_assistive_subsidies}. Additionally, organizations like the WHO advocate for affordable AI tools to address accessibility gaps \cite{who_accessibility_report}.

\begin{itemize}
    \item Governments should prioritize funding for affordable assistive technologies \cite{norway_assistive_subsidies}.
    \item Policies should encourage research and development in accessible AI tools \cite{accessible_ai_policy}.
    \item NGOs such as the National Federation of the Blind provide training and support to visually impaired individuals \cite{ngo_involvement_accessibility}.
\end{itemize}


\section{Challenges}
\label{sec:challenges}

While AI-powered assistive technologies have revolutionized accessibility for visually impaired individuals, several challenges remain. Addressing these issues is crucial for ensuring equitable access, user satisfaction, and sustained innovation.

\subsection{Affordability and Accessibility}

Many AI-powered assistive technologies are prohibitively expensive for individuals in low-income settings. For instance, devices like OrCam MyEye and Envision Glasses cost upwards of \$2,000, making them inaccessible to a significant portion of the visually impaired population \cite{orcam_2023, envision_glasses_review}. Additionally, the availability of these tools is often limited in rural or underserved regions, further exacerbating disparities.

The "digital divide" is another critical challenge, as reliable internet connectivity, necessary for many AI applications, remains unavailable to millions globally. This lack of infrastructure further marginalizes individuals in remote or under-resourced areas.

\subsection{Ethical and Privacy Concerns}

AI systems that process sensitive information, such as facial recognition tools or text-to-speech applications, raise ethical concerns. These include:
\begin{itemize}
    \item \textbf{Data Privacy:} Many applications require continuous internet connectivity, leading to concerns about the collection and storage of personal data \cite{privacy_ai_data_33}.
    \item \textbf{Algorithmic Bias:} Some AI models exhibit biases that may disadvantage specific user groups, such as non-native speakers or individuals with unique facial features \cite{algorithmic_bias_facial_recognition_37}.
    \item \textbf{Potential Misuse:} Facial recognition tools may be misused for unauthorized surveillance or identification, posing risks to individual freedom and security \cite{misuse_of_facial_recognition_16}.
\end{itemize}

\subsection{Usability and User Experience}

Although these technologies are advanced, their effectiveness depends on intuitive design and user-friendly interfaces. Some users report difficulty in navigating complex interfaces or adapting to multi-functional tools \cite{usability_review}. Moreover, many systems rely heavily on high-speed internet and frequent updates, which can be challenging in low-resource environments.

\subsection{Limited Cultural and Linguistic Adaptation}

Many AI tools are optimized for English-speaking users, with limited support for other languages or cultural nuances. For instance, voice assistants may struggle with diverse accents or fail to integrate region-specific commands, reducing their effectiveness in multilingual societies \cite{language_adaptation_review}.

\section{Future Directions}
\label{sec:future_directions}

\subsection{Affordable Solutions Through Partnerships and Open-Source Tools}

Collaborations between governments, non-profits, and tech companies can help subsidize the cost of assistive technologies. Programs like Microsoft's AI for Accessibility initiative provide funding for projects aimed at creating low-cost solutions \cite{ai_accessibility_initiative}. Expanding these initiatives globally could improve accessibility for underserved populations.

Open-source software and DIY (Do-It-Yourself) assistive technologies also offer affordable and customizable alternatives. For example, projects like NVDA (NonVisual Desktop Access) demonstrate how free, community-driven solutions can effectively bridge accessibility gaps \cite{nv_access_nvda}.

\subsection{Advancements in Ethical AI}

To address privacy and bias concerns, future technologies must incorporate:
\begin{itemize}
    \item \textbf{Federated Learning:} AI systems that process data locally on devices rather than relying on cloud servers offer enhanced privacy by minimizing the transfer of sensitive information \cite{myakala2024federated}.
    \item \textbf{Bias Mitigation:} It is crucial for developers to prioritize the use of diverse and inclusive datasets during the training of AI models to effectively minimize algorithmic biases and ensure fairness across different demographic groups \cite{inclusive_datasets_bias_mitigation}.
    \item \textbf{Regulatory Oversight:} Policymakers must establish robust guidelines to prevent the misuse of sensitive technologies, such as facial recognition, which has raised concerns about surveillance, bias, and wrongful use. Effective regulations should prioritize transparency, data protection, and accountability in AI development \cite{almeida2021ethics, gao2021facial, openthegov2019guide}.

\end{itemize}

\subsection{Enhanced Usability, Localization, and Personalization}

Future designs must focus on user-centric approaches to simplify interfaces and reduce cognitive load. Additionally, expanding linguistic and cultural support will ensure that tools are effective in diverse regions. For example, training NLP models on low-resource languages could make voice assistants more inclusive \cite{nlp_low_resource}.

Personalization is another key area of improvement. Allowing users to customize tools according to their specific needs and preferences, such as adjustable voice speed in screen readers or personalized object recognition databases, can significantly enhance user satisfaction.

\subsection{Integration with Emerging Technologies}

Integrating assistive tools with emerging technologies like augmented reality (AR), the Internet of Things (IoT), and 5G networks could unlock new possibilities. For instance:
\begin{itemize}
    \item \textbf{Augmented Reality:} AR-enabled smart glasses could provide enhanced scene analysis with overlaid audio descriptions .
    \item \textbf{IoT Integration:} Connecting assistive devices with smart homes or wearables could streamline daily activities.
    \item \textbf{5G Networks:} Faster connectivity would enable real-time processing, reducing delays in applications like navigation or object detection.
\end{itemize}

\subsection{Fostering Interdisciplinary Collaboration and Participatory Design}

Collaboration between researchers, developers, and end-users is essential to ensure that assistive technologies meet practical needs. Participatory design approaches that actively involve visually impaired individuals during development can lead to more effective and user-friendly solutions. For instance, user feedback on prototype designs can identify challenges early in the development cycle, improving overall usability and adoption.

\section{Conclusion}
\label{sec:conclusion}

AI-powered assistive technologies represent a significant advancement in improving the quality of life for visually impaired individuals. By leveraging cutting-edge innovations such as computer vision, natural language processing, and wearable devices, these tools have enabled greater independence, accessibility, and inclusion in various aspects of life, including education, mobility, and social interaction.

This review highlighted the applications and impact of these technologies, emphasizing their ability to address challenges in independent living, navigation, and education. Tools such as smart glasses, screen readers, and navigation aids have empowered visually impaired individuals to interact with their environments in ways that were previously impossible. Moreover, the role of AI in enhancing social interactions and mental well-being underscores its holistic value.

Despite their transformative potential, several challenges persist. Issues of affordability, accessibility, and ethical concerns remain critical barriers to widespread adoption. Additionally, the digital divide and limited cultural and linguistic adaptation must be addressed to ensure that these technologies are inclusive and equitable.

Looking ahead, the future of AI-powered assistive technologies lies in fostering interdisciplinary collaboration, advancing ethical AI practices, and integrating emerging technologies. For instance, AI systems integrated with smart home environments could provide contextual information and automate daily tasks, while personalized virtual assistants could learn user preferences and anticipate individual needs. Such innovations have the potential to further enhance autonomy and efficiency for visually impaired users.

To realize this vision, collaboration among researchers, developers, policymakers, and disability advocates is essential. Researchers must focus on creating affordable and adaptable solutions, while policymakers should implement supportive frameworks that encourage accessibility and inclusivity. Open-source initiatives and participatory design processes, involving visually impaired individuals at every stage, will be critical to ensuring that future technologies are both innovative and user-centric.

By addressing these challenges and embracing future opportunities, AI-powered assistive technologies can continue to bridge accessibility gaps, promote independence, and foster an inclusive society where individuals with visual impairments can thrive. Now is the time for collective action to prioritize the development and equitable distribution of these transformative tools, ensuring that no one is left behind.

\bibliographystyle{unsrt}  
\bibliography{references}

\end{document}